\documentclass{article}
\usepackage[spanish]{babel}
\usepackage{arxiv}

\usepackage[utf8]{inputenc} 
\usepackage[T1]{fontenc}    
\usepackage{hyperref}       
\usepackage{url}            
\usepackage{amsfonts}       
\usepackage{nicefrac}       
\usepackage{pdflscape}      
\usepackage{microtype}      

\usepackage{graphicx}
\usepackage{natbib}
\usepackage{doi}
\usepackage{amsmath}

\usepackage{tabularx} 
\usepackage[labelfont=bf]{caption}
\usepackage{booktabs} 

\usepackage{xcolor}
\usepackage{subfig}
\usepackage{wrapfig}

\usepackage{multirow} 
\usepackage{float}

\addto\captionsspanish{
  
}

\title{Automatización de Informes Geotécnicos para Macizos Rocosos con IA}

\author{ 
\href{https://orcid.org/0009-0006-0456-8287}{\includegraphics[scale=0.1]{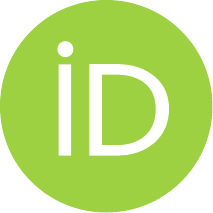}\hspace{1mm}Christofer Valencia}
\And
\href{}
{\includegraphics[scale=0.1]{orcid.pdf}\hspace{1mm}Alexis Llumigusín}
\And
\href{https://orcid.org/0009-0002-7999-9197}{\includegraphics[scale=0.1]{orcid.pdf}\hspace{1mm}Silvia Alvarez}
\And
\href{https://orcid.org/0009-0003-8760-7453} {\includegraphics[scale=0.1]{orcid.pdf}\hspace{1mm}Abrahan Arias}
\And
\href{https://orcid.org/0000-0001-6715-191X}
{\includegraphics[scale=0.1]{orcid.pdf}\hspace{1mm}Christian Mejia-Escobar}
\\
\\
\texttt{\{covalencia,syalvarezm,adllumigusin,ajariasl,cimejia\}@uce.edu.ec}
\\
Carrera de Geología, Universidad Central del Ecuador, Quito, 170129 \\
}

\renewcommand{\shorttitle}{\textit{arXiv} Template}

\hypersetup{
pdftitle={A template for the arxiv style},
pdfsubject={q-bio.NC, q-bio.QM},
pdfauthor={David S.~Hippocampus, Elias D.~Striatum},
pdfkeywords={First keyword, Second keyword, More},
}

\begin{document}
\maketitle


\begin{abstract}
Los informes geotécnicos son cruciales para evaluar la estabilidad de las formaciones rocosas y garantizar la seguridad en la ingeniería moderna.
Tradicionalmente, estos informes se elaboran de manera manual a partir de observaciones en campo con brújulas, lupas y libretas.
Este método resulta lento, propenso a errores y subjetivo en las interpretaciones.
Para superar estas limitaciones, se propone el uso de técnicas de inteligencia artificial para la generación automática de informes mediante el procesamiento de imágenes y datos de campo.
La metodología se basó en recopilar fotografías de afloramientos rocosos y muestras de mano con sus respectivas descripciones, así como informes realizados en la cátedra de Geotecnia.
Estos recursos se emplearon para definir el esquema del informe, realizar la ingeniería del prompt y validar las respuestas de un modelo de lenguaje grande multimodal (MLLM).
El refinamiento iterativo de prompts hasta obtener instrucciones estructuradas y específicas para cada sección del informe demostró ser una alternativa eficaz al costoso proceso de ajuste fino del MLLM.
La evaluación del sistema establece valores de 0.455 y 0.653 para las métricas BLEU y ROUGE-L, respectivamente, lo que sugiere descripciones automáticas comparables a las realizadas por expertos.
Esta herramienta, accesible mediante la web, con una interfaz intuitiva y capacidad de exportación a formatos estandarizados, representa una innovación y contribución importante para profesionales y estudiantes de la geología de campo.

\end{abstract}


\section{Introducción}
\label{sec:intro}
El producto final de la exploración geológica es el informe de reconocimiento de campo. 
Específicamente, los informes geotécnicos de campo para formaciones rocosas documentan las propiedades geomecánicas de estas estructuras, así como su complejidad y estabilidad.
Por ende, se convierten en un insumo clave para la toma de decisiones en ingeniería, la identificación de riesgos geológicos y el cumplimiento de normas técnicas y legales \cite{perez2010}.


Tradicionalmente, la elaboración de este tipo de informe es un proceso que comienza con la toma de datos en campo, utilizando herramientas como la brújula, la libreta, la lupa y el martillo geológico. Posteriormente, se procede a la interpretación de estos datos y a la redacción del documento final, etapas cuyo desarrollo depende en gran medida del criterio del especialista.


La naturaleza manual de estas actividades las convierte en un proceso que no solo consume mucho tiempo, sino también en ineficiente, repetitivo y subjetivo \cite{aoxue2023}.
Aunque existen plantillas de informes disponibles, estas son de alta complejidad y completarlas manualmente es una fuente potencial de errores, lo que disminuye la eficiencia y reduce la calidad del producto final \cite{lei2020research}.

En contraste, la Inteligencia Artificial (IA) ofrece ventajas como la capacidad de automatizar tareas repetitivas, procesar grandes cantidades de datos, aumentar la objetividad, la eficiencia y la precisión \cite{DelaTorre2023artificial}\cite{baghbani2022}.
Si bien la IA ha sido aplicada con éxito para la generación de informes técnicos en varias disciplinas, desde la medicina hasta la ingeniería \cite{Yagamurthy2023natural}\cite{Choi2025apalancamiento}, en geología aún no se han desarrollado investigaciones enfocadas en automatizar la elaboración de informes de campo específicos para geotecnia de macizos rocosos, reducir el tiempo de entrega y promover la estandarización de estos informes \cite{Martinez2024Geoparques}.

Wang et al. (2022) \cite{Wang2022visual} crean un marco computacional para analizar informes geológicos de exploración minera mediante procesamiento de lenguaje natural (NLP) y visualización. Utilizaron un algoritmo TF-IDF mejorado con una longitud de palabra para extraer palabras clave, logrando mayor precisión en textos chinos (F1-score del 77.14\%). El enfoque facilita la interpretación rápida de datos complejos.
    
Dimeski \& Rahimi (2022) \cite{Dimeski2022automatic} desarrollaron un sistema para extraer datos estructurados de perforación mineral de informes no estructurados, utilizando modelos Bi-LSTM-CRF y BERT. Con un dataset de 23 compañías mineras australianas, BERT superó al Bi-LSTM-CRF (F1-score del 87\% vs. 77\%), destacando en generalización entre materiales y empresas. El estudio subraya el potencial de NLP para optimizar permisos mineros y reducir costos manuales, aunque señala desafíos en la consistencia de formatos y terminología.

Lopez \& Al-Dossari (2024) \cite{lopez2024automated} presentan un sistema automático para estudiar filtraciones en obras geotécnicas mediante NLP y deep learning. Usaron un modelo que combina redes neuronales (CNN-LSTM) con atención para predecir datos como la presión del agua y el flujo. Crearon una herramienta que extrae información directamente de informes técnicos, incluso si están desordenados. Este enfoque hace el análisis 200 veces más rápido que a mano y ayuda a detectar riesgos en infraestructuras de forma más rápida y confiable.

Los trabajos anteriores reconocen que la redacción automatizada todavía presenta limitaciones debido a su enfoque en tareas específicas como la extracción de palabras clave, la clasificación litológica o la generación parcial de texto.
Este proyecto supera esta barrera adaptando al ámbito geológico la IA moderna, caracterizada por sus capacidades generativas y multimodales, sin necesidad de reentrenar modelos.
Un cuidadoso diseño de prompts permite la redacción técnica de informes geotécnicos completos, coherentes y precisos directamente a partir de imágenes y datos de campo ingresados por el usuario.
A través de una interfaz web pública y fácil de usar, la solución desarrollada con herramientas de software libre y validada con las métricas BLEU y ROUGE-L, ofrece una automatización novedosa diseñada para mejorar la productividad en las actividades técnicas de campo y oficina.

Nuestro propósito es utilizar una estrategia de IA para generar informes geotécnicos de manera automatizada combinando técnicas de Visión por Computadora (CV) y Procesamiento de Lenguaje Natural (NLP).
Anteriormente, para combinar CV y NLP, era necesario integrar múltiples modelos especializados; sin embargo, los recientes modelos de lenguaje grandes multimodales (MLLM) simplifican este proceso por ser un único modelo que fusiona estas técnicas.
La solución propuesta se basa en un MLLM para procesar y entender tanto el lenguaje (texto) como el contenido visual (imágenes) de forma simultánea.

La contribución principal de este trabajo es el desarrollo de una aplicación web gratuita y disponible en línea con una interfaz de usuario sencilla para el ingreso de los datos, tanto texto como imágenes, y crear informes de manera más rápida y precisa.
Esta herramienta no solo aumenta la eficiencia de profesionales y estudiantes, sino que les permite concentrarse en tareas de mayor valor, como la interpretación de resultados basada en el contexto, lo que es esencial para lograr un análisis profundo y preciso \cite{paucar2024}.
Además de incrementar la calidad de la información en el ámbito geológico-geotécnico, esta innovación también podría facilitar la estandarización de los informes de campo, mejorando la comprensión y análisis de la información geológica \cite{Rodriguez2025IA}.

El contenido del documento se presenta de la siguiente manera: un contexto general del proyecto y los trabajos relacionados son abordados en la Sección \ref{sec:intro}. La Sección \ref{sec:mehodology} detalla las diferentes actividades de la metodología empleada. La Sección \ref{sec:experiments} describe la parte experimental, la evaluación del modelo y analiza los resultados. El desarrollo de la aplicación web se expone en la Sección \ref{sec:app}. Finalmente, la Sección \ref{sec:conclusion} enuncia las conclusiones de este trabajo y posibles líneas de investigación a futuro.

\section{Metodología}
\label{sec:mehodology}
Nuestro objetivo es automatizar la generación de informes geotécnicos para macizos rocosos mediante un MLLM aprovechando su capacidad de procesar texto e imagen de forma unificada.
Así, se logra optimizar el tiempo de entrega y elevar la calidad de los informes, permitiendo que tanto profesionales como estudiantes se concentren plenamente en la interpretación de los resultados.
La Figura \ref{fig:flujograma} presenta la metodología desarrollada, que consta de siete etapas; cada una de estas etapas es detallada a continuación:

\begin{figure}[!htb]
    \centering
    \includegraphics[width=\linewidth]{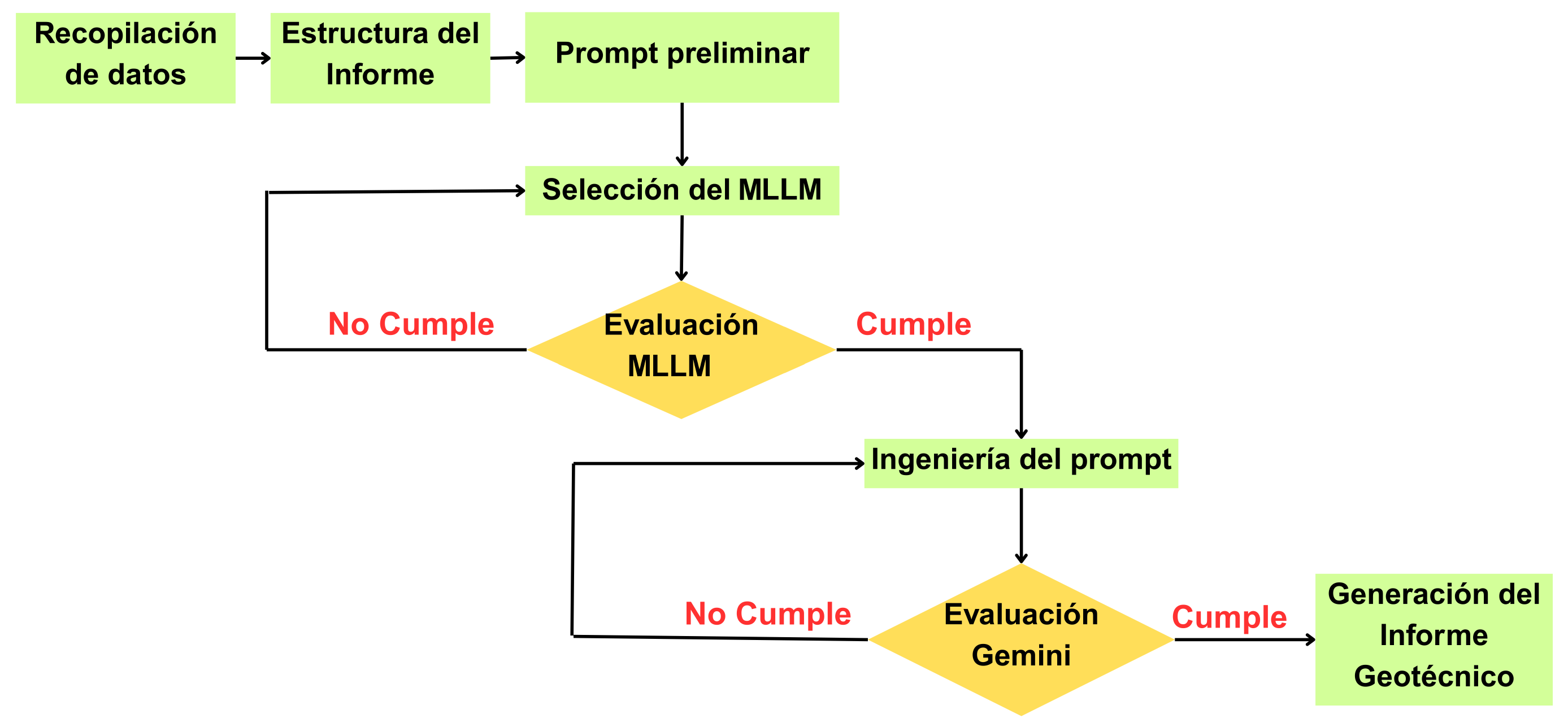}
    \caption{Diagrama de flujo de la metodología seguida para la generación automática del informe geotécnico.}
    \label{fig:flujograma}
\end{figure}

\subsection{Recopilación de datos} 
\begin{figure}[!htb]
    \centering
    \includegraphics[width=0.9\linewidth]{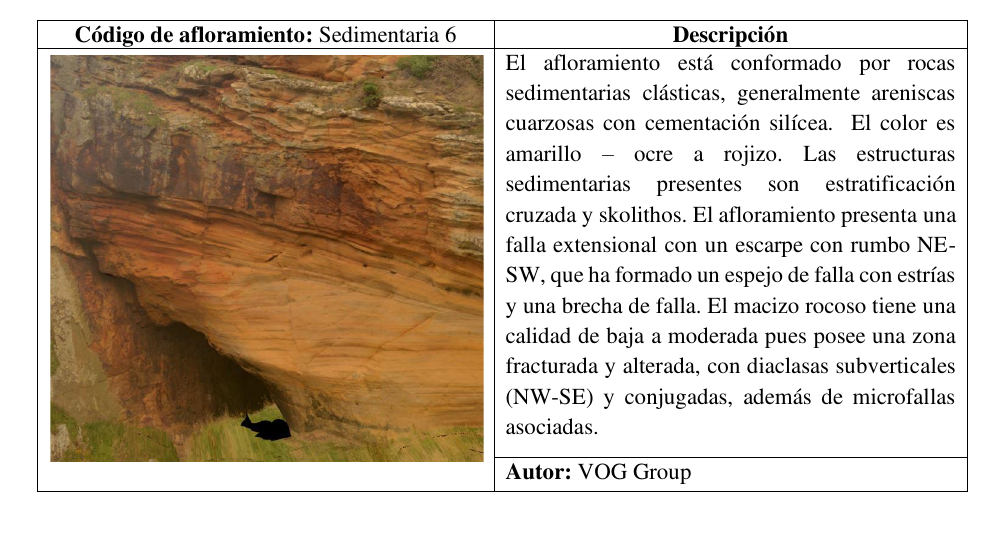}
    \caption{Ejemplo de descripción geológica-geotécnica de campo de una imagen de afloramiento.}
    \label{fig:des_campo}
\end{figure}

Los datos son la materia prima de una solución basada en IA. En este caso, los datos consisten en imágenes y texto.
Se recopilaron un total de 30 imágenes de afloramientos de macizos rocosos descargadas del sitio \textit{V3Geo}\footnote{\url{https://v3geo.com/}}, un repositorio público de modelos virtuales en 3D \cite{gc-5-67-2022}. Estas imágenes se acompañan de su respectiva caracterización textual de autores como \cite{Ingdahl1989}, \cite{Tricart2006},  \cite{Young2009}, \cite{Faerseth2011}, \cite{Malavieille2015}, entre otros. Un ejemplo de estos datos se presenta en la Figura \ref{fig:des_campo}.

Las imágenes tienen un formato JPG con una resolución de textura de 4096 x 4096 píxeles y un peso de alrededor de 1 MB, esto con el fin de tener una mejor compatibilidad con diversas plataformas y navegadores web. La caracterización textual de los afloramientos fue tabulada en el software Microsoft Excel. 

\begin{figure}[!htb]
    \centering
    \includegraphics[width=0.3\linewidth]{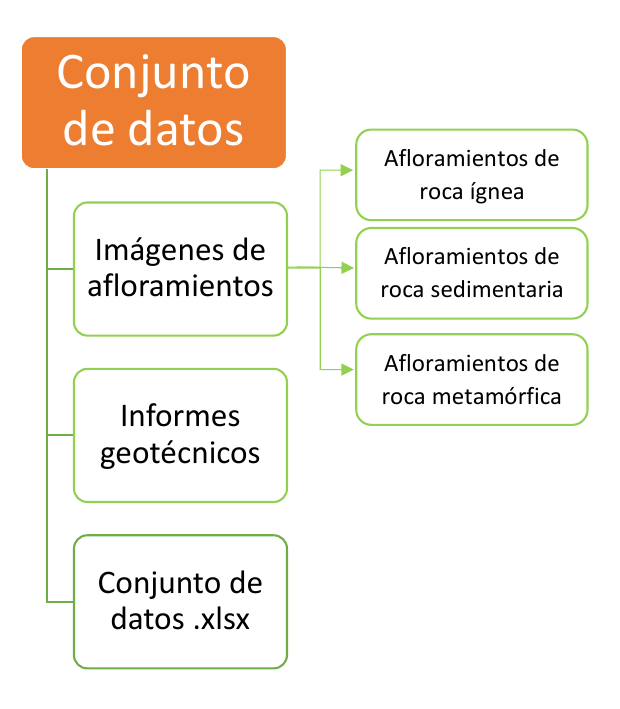}
    \caption{Diagrama de la estructura del conjunto de datos.}
    \label{fig:ejem_dataset}
\end{figure}

El conjunto de datos (\textit{dataset}) fue almacenado de forma local y en \textit{Google Drive}. Se encuentra disponible al público mediante la dirección URL: \href{https://drive.google.com/drive/folders/1rVjlQ4uWzEU54AuiqQipIDsEDjrOjhBK}{\textcolor{blue}{dataset}} y su estructura se ilustra en la Figura \ref{fig:ejem_dataset}.

La carpeta principal contiene dos subcarpetas y el archivo de Excel. Una subcarpeta es para las 30 imágenes de afloramientos, con 10 imágenes para cada una de las tres categorías según su origen petrogenético. Este conjunto está balanceado y define un código de identificación con el tipo de roca y número de afloramiento, por ejemplo, ``Sedimentaria 1''.
El archivo XLSX contiene las descripciones de cada afloramiento descompuestas en sus componentes: ID, tipo de roca, geología, color predominante, estructuras principales, calidad del macizo rocoso y descripción de juntas.
Ambos insumos serán utilizados para evaluar el desempeño del MLLM.
La subcarpeta restante almacena los informes geotécnicos recopilados, los cuales son útiles para establecer la estructura del informe final. 

\subsection{Estructura del informe}
\begin{figure}[H]
\centering
    \includegraphics[width=0.35\linewidth]{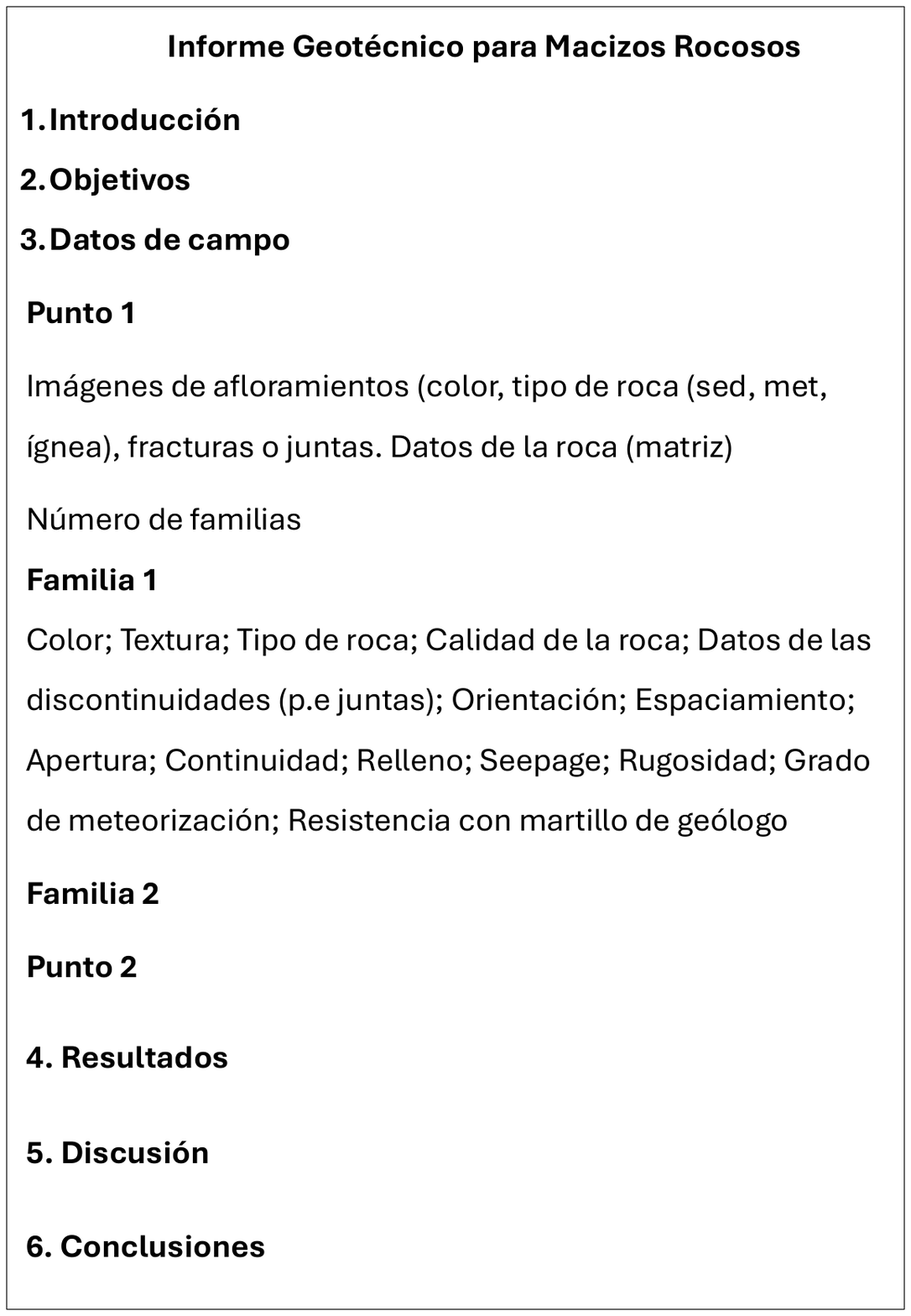}
    \caption{Esquema general de un informe geotécnico de campo para roca.}
    \label{fig:estru_infor}
\end{figure}

Es necesario determinar las diferentes partes que conformarán el informe geotécnico de campo para macizos rocosos.
Para este propósito, se consideraron los informes en formato PDF recopilados de la cátedra de Mecánica de Rocas y Geotecnia de la Universidad Central del Ecuador, los cuales se encuentran disponibles en la dirección URL: \href{https://drive.google.com/drive/folders/10sbBFU4RyBfSVuslyKjLD0zJOiGA6FnG?usp=drive_link}{\textcolor{blue}{informes de campo}}.
La Figura \ref{fig:estru_infor} ilustra el contenido general de este tipo de informes.

Mediante el análisis y la comparación de estos informes, los cuales siguen un esquema estándar sobre salidas de campo, se estableció la estructura descrita en la Tabla \ref{tb-1} .

\renewcommand{\arraystretch}{1.5}
\begin{table}[H]
\centering
\caption{Estructura del informe geotécnico de formaciones rocosas.}
\label{tb-1}
\scalebox{0.9}{
\begin{tabular}{|>{\bfseries}l|l|p{8.5cm}|}
\hline
\textbf{Sección} & \textbf{Subsecciones} & \textbf{Descripción} \\
\hline
\textbf{Portada} & - & Presenta datos de universidad, facultad, carrera, nombre del informe, nombre del proyecto, asignatura, autor(es) y fecha. \\
\hline
1. \textbf{Introducción} & - & Describe el tema del informe y la ubicación del proyecto. \\
\hline
2. \textbf{Objetivos} & 2.1 Objetivo general & Es generado a partir del tema del proyecto. \\
\cline{2-3}
& 2.2 Objetivos específicos & Se generan dos objetivos a partir del tema del proyecto. \\
\hline
3. \textbf{Datos de campo} & 3.1 Tabla de afloramientos & En una tabla se registra cada afloramiento con un código numérico, las coordenadas (x,y,z) y el sistema de referencia. \\
\cline{2-3}
& 3.2 Tabla de características de rocas & En una tabla se registran: el tipo de roca, la matriz, textura, mineralogía, tamaño de grano y calidad del macizo. \\
\cline{2-3}
& 3.3 Afloramiento & Para cada afloramiento se presenta su imagen, de la muestra de mano y las gráficas de barras y estereogramas según sus características geotécnicas. \\
\hline
4. \textbf{Resultados} & - & Descripciones generadas de manera automática con base en la imágenes y datos proporcionados por el usuario. \\
\hline
5. \textbf{Discusión} & - & Se conforma con los resultados obtenidos, una interpretación técnica de los patrones estructurales observados y su influencia en la calidad del macizo rocoso. \\
\hline
6. \textbf{Conclusiones} & - & Sintetiza los aspectos analizados, como la relación entre las discontinuidades y la calidad del macizo rocoso. \\
\hline
\textbf{Anexos} & Anexo A & Incluye la tabla de clasificación RMR por afloramiento con los parámetros geotécnicos (número de familias de juntas, orientación de juntas, UCS, RQD, espaciamiento, continuidad, apertura, rugosidad, alteración, agua freática) ingresados por el usuario. Estos son puntuados, obteniendo así un total dentro de la clasificación RMR. \\
\cline{2-3}
& Anexo B & Presenta los resultados de la estadística por afloramiento, el número de familias de juntas y el RMR máximo y mínimo. Además de la tabla de clasificación de macizos rocosos RMR. \\
\hline
\end{tabular}
}
\end{table}

La estructura del informe consideró además los parámetros sugeridos por las organizaciones ISRM\footnote{International Society for Rock Mechanics and Rock Engineering (\url{https://isrm.net/})} \cite{isrm1981} y ASTM \footnote{American Society for Testing and Materials (\url{https://www.astm.org/})} \cite{astm2019} para la caracterización de macizos rocosos.
Este es el caso de la resistencia a la compresión simple de una roca mediante el martillo de Schmidt \cite{astm2025}\cite{aydin2008}, cuyo cálculo es importante para establecer una clasificación para macizos rocosos como el RMR \cite{bieniawski1989}.
A partir de la estructura obtenida, se diseñan los \textit{prompts}, que son instrucciones detalladas que guían al MLLM para generar de forma automática y precisa el contenido de cada sección del informe.

\subsection{Prompt preliminar}

Para seleccionar el MLLM más adecuado, se creó un prompt de prueba que permitió comparar el rendimiento de cada modelo.
Hay que recordar que el MLLM constituye el componente central o cerebro del sistema de generación automática de informes.
Para una evaluación justa y precisa del modelo, es esencial que el prompt esté bien formulado, ya que la calidad de la respuesta del modelo está directamente ligada a la calidad de la instrucción que recibe.

\begin{figure}[H]
    \centering
    \includegraphics[width=0.8\linewidth]{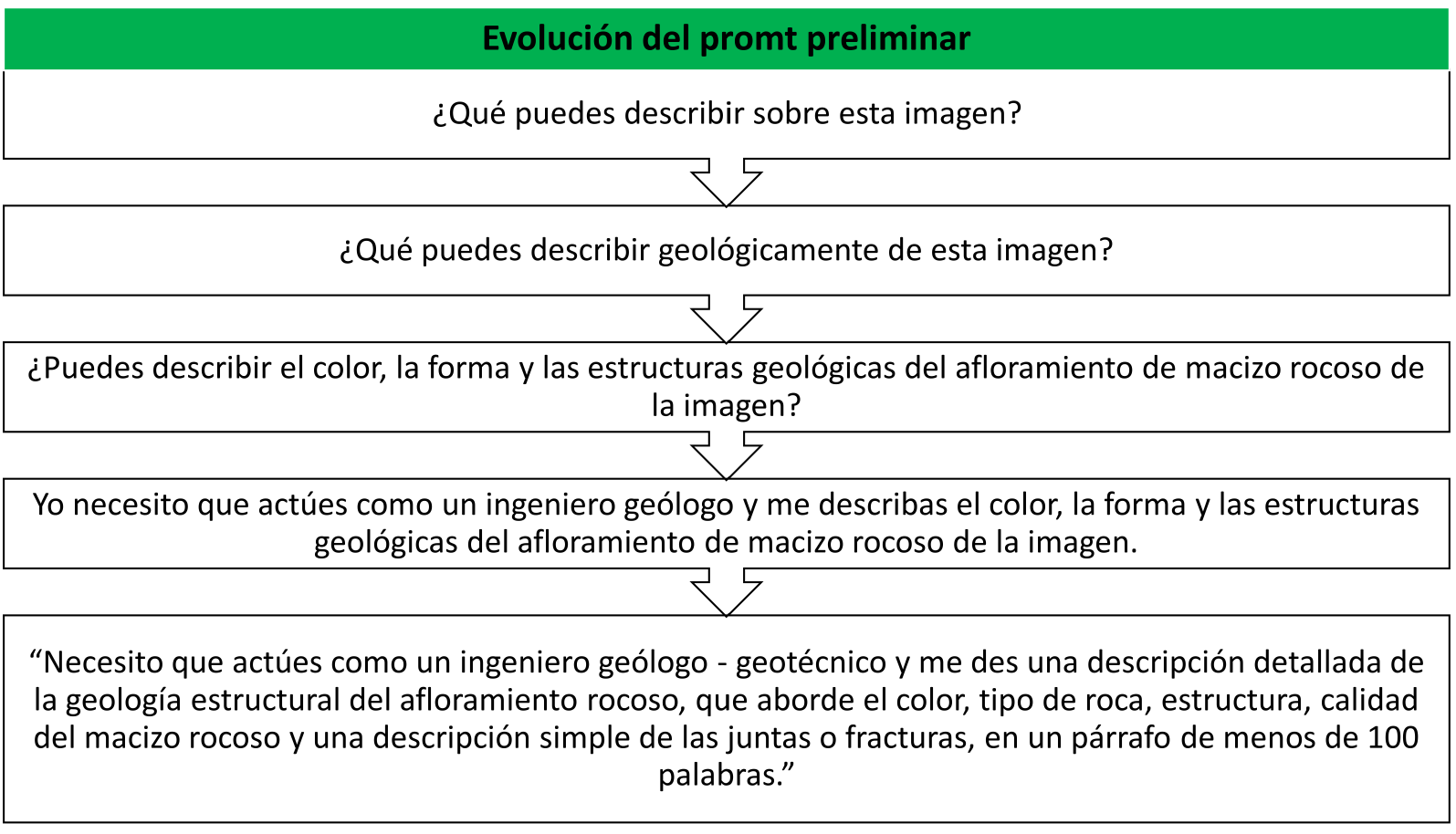}
    \caption{Evolución de la complejidad y detalle en la redacción del prompt preliminar.}
    \label{fig:evolución}
\end{figure}

A través de una serie de modificaciones progresivas, el prompt se fue perfeccionando de forma iterativa, desde una instrucción inicial simple hasta una versión más técnica y elaborada. La Figura \ref{fig:evolución} ilustra esta evolución.

La estructura del prompt que permitió al MLLM generar una interpretación técnica y coherente de las imágenes de afloramientos integra los siguientes componentes:

\begin{center}
\small\texttt{Prompt preliminar = [Rol de experto] + [Petición de descripción de la imagen] + [Características geológicas requeridas] + [Características geotécnicas requeridas] + [Extensión de la descripción]}    
\end{center}

El componente de las características geológicas incluye el color, el tipo de roca (ígnea, sedimentaria o metamórfica) y las estructuras geológicas.
Las características geotécnicas corresponden a la calidad del macizo rocoso y la descripción de juntas y/o discontinuidades.
Ambos tipos de características son básicas para la descripción de las imágenes y necesarias para rellenar la sección de resultados.
Por último, la extensión del prompt fue limitada a 100 palabras debido a que el MLLM puede extenderse demasiado y divagar en la descripción generada.

\subsection{Selección del MLLM}

El prompt preliminar fue útil para interactuar con los MLLMs disponibles en la web y obtener descripciones geotécnicas precisas a partir de imágenes de afloramientos.
Además de la disponibilidad, los MLLMs se eligieron por factibilidad de uso, viabilidad técnica y accesibilidad en el momento del estudio, incluyendo modelos que ofrecen APIs y otros de código abierto que podían ser implementados con los recursos disponibles del proyecto.
Estos modelos se despliegan en la Tabla \ref{tb1} con algunas de sus especificaciones principales.

\begin{table}[H]
\centering
\caption{Comparación de los modelos multimodales considerados para el estudio.}
\label{tb1}
\scalebox{0.85}{
\begin{tabular}{ccccc}
\toprule
\textbf{Modelo} & \textbf{Tipo de dato} & \textbf{Memoria (FP16)} & \textbf{Accesibilidad} & \textbf{Costo API/Cloud} \\
\midrule
Phi-4-MM & Texto e imagen & 8--20 GB & Limitada (Microsoft) & Posible pago en Azure \\
Phi-3-V & Texto y visión & 14 GB / 28 GB & Abierto (Hugging Face) & Gratis (local) \\
LLaVA-1.5 & Texto e imagen & 14 GB & Abierto & Gratis (local) \\
Gemini 1.5 Flash & Texto, imagen y video & $\sim$70 GB & API (Google AI Studio) & Gratis (60 intentos por API key) \\
Copilot & Basado en GPT-4/GPT-3.5 & 350 GB / 3.6 TB & Integrado (Microsoft) & \$10--\$30/mes (suscripción) \\
\bottomrule
\end{tabular}
}
\end{table}

La estrategia de evaluación consistió en probar el prompt preliminar con cada MLLM usando tres imágenes de afloramientos rocosos, una por cada clasificación petrogenética: ígnea, sedimentaria y metamórfica.
Las descripciones de cada modelo se tabularon en Excel para posteriormente analizar y comparar las respuestas. Un ejemplo de respuestas para una imagen de afloramiento dada se despliega en la Tabla \ref{tab_compa_res}.

\begin{table}[H]
\centering
\caption{Comparación de las respuestas proporcionadas por los MLLMs.}
\label{tab_compa_res}
\scalebox{0.95}{
\begin{tabular}{p{3cm}p{3cm}p{3cm}p{3cm}p{3cm}}
\toprule
\textbf{Características} & \textbf{LLaVA} & \textbf{Copilot} & \textbf{Gemini 1.5 Flash} & \textbf{Phi-3} \\
\midrule
\textbf{Geológicas (color, tipo de roca y estructura)} & 
Identifica el tipo de roca por el origen petrogenético (ígnea, sedimentaria y metamórfica). En ocasiones no reconoce bien el color. No identifica estructuras geológicas más allá de juntas. & 
Identifica la clasificación de la roca (arenisca, basalto, gneis). Reconoce los colores del afloramiento. Reconoce estructuras geológicas presentes en ambientes ígneos, sedimentarios y metamórficos. & 
Identifica tanto el tipo de roca por el origen petrogenético como su clasificación. Reconoce bien los colores del afloramiento. Reconoce estructuras geológicas presentes en ambientes ígneos, sedimentarios y metamórficos. & 
No es precisa para identificar el tipo de roca. En ocasiones no reconoce bien el color. El reconocimiento de estructuras geológicas es limitado. \\
\midrule
\textbf{Geotécnicas (calidad del macizo rocoso y descripción de juntas o discontinuidades)} & 
Identifica la presencia de juntas. La interpretación del origen de las juntas es en ocasiones erróneo. & 
Identifica la presencia de juntas y su orientación vertical y subvertical. La interpretación del origen de las juntas es más verosímil. & 
Identifica la presencia de juntas y su orientación vertical, subvertical y horizontales. La interpretación del origen de las juntas es más verosímil. & 
Identifica la presencia de juntas. La interpretación del origen de las juntas es en ocasiones erróneo. \\
\bottomrule
\end{tabular}
}
\end{table}

Las diferentes respuestas obtenidas fueron comparadas para determinar su precisión y coherencia.
Los resultados demuestran diferencias significativas en la capacidad de los modelos evaluados para describir características geológicas y geotécnicas de las imágenes de afloramientos rocosos.
Gemini 1.5 Flash sobresale como el modelo más completo, al identificar con precisión tanto el origen petrogenético como la clasificación de rocas, reconocer colores y estructuras geológicas en diversos ambientes (ígneos, sedimentarios y metamórficos), y ofrecer interpretaciones verosímiles sobre discontinuidades.
Su capacidad para integrar múltiples variables lo hace ideal para aplicaciones técnicas avanzadas.

En el caso de Copilot muestra un buen rendimiento, especialmente en la clasificación de rocas y el reconocimiento de estructuras. Su interpretación de juntas y discontinuidades es destacable, lo que sugiere utilidad en contextos geotécnicos básicos. Por otro lado, LLaVA presenta limitaciones notables; aunque identifica tipos de roca por origen, falla en reconocer colores y estructuras complejas, y sus interpretaciones sobre juntas son menos confiables. Phi-3 resulta el menos preciso, con dificultades para identificar tipos de roca, colores y estructuras, lo que limita su aplicabilidad en geología. Su bajo rendimiento en el análisis de discontinuidades refuerza la necesidad de mejorar su entrenamiento con datos geotécnicos especializados.

Finalmente, el modelo elegido fue Gemini 1.5 Flash por ser un modelo multimodal que para acceder a él se necesita una API key que permite 60 intentos gratis, además de ser el modelo con las descripciones de imágenes más coherentes y de mejor calidad. Otros MLLMs gratuitos no fueron considerados debido a su baja calidad de redacción técnica para el informe.

\section{Experimentación y resultados} 
\label{sec:experiments}

Tras definir la estructura del informe geotécnico y elegir el modelo Gemini 1.5 Flash, se procede a la experimentación.
Tradicionalmente, esta etapa documenta el entrenamiento, ajuste y evaluación del modelo.
En un inicio, nuestra investigación consideró el reentrenamiento de los modelos LLaVA y Phi-3, con el propósito de mejorar su desempeño en la generación de descripciones técnicas para afloramientos de macizos rocosos, utilizando un conjunto de datos propio.
Sin embargo, las limitaciones de memoria RAM y GPU en la versión gratuita de Google Colab y en los equipos locales imposibilitaron la realización del proceso de ajuste fino sobre los modelos.

Alternativamente, se opta por un cambio de paradigma: en lugar de ``enseñar'' al modelo, se procede a ``instruirlo''.
En este caso, el MLLM ya no es visto como un cerebro incompleto que necesita ser reentrenado para una tarea específica; se le ve como un experto con un conocimiento vasto, pero sin dirección.
El prompt se convierte en la herramienta para dar esa dirección, habilitando esta capacidad del modelo sin la necesidad de modificar su arquitectura interna.

Por lo tanto, la parte experimental del proyecto se centra en un proceso de ingeniería de prompt, probando cómo una serie de instrucciones específicas cuidadosamente diseñadas y refinadas permiten alcanzar los resultados deseados, es decir, generar con éxito cada una de las secciones que conforman el informe geotécnico de macizos rocosos.

\subsection{Ingeniería del prompt}

La ingeniería del prompt fue una fase clave del proyecto. Debido a las múltiples secciones que conforman el informe, se aplicó el principio de que un gran problema se puede dividir en subproblemas más pequeños, o el conocido ``divide y vencerás''.
La estrategia consiste en crear un prompt particular para cada sección del informe: introducción, descripción del afloramiento, descripción de la muestra de mano, resultados, discusión y conclusiones.
La combinación de estos prompts individuales produce un gran prompt final que representa todo el informe geotécnico.

\begin{center}
\small
\texttt{PromptFinal = PromptObjetivos + PromptIntroducción + PromptDescripciónAfloramiento + PromptMuestraDeMano + PromptEsclerómetro + PromptDiscusión + PromptConclusiones}    
\end{center}

La estructuración de los prompts permite que el modelo procese la información de manera adecuada y la organice en base a las secciones que componen el informe. 
En principio, se diseñó el prompt para la descripción de imágenes de afloramientos, que constituye el eje central del informe.
Una vez perfeccionado este prompt, se continúa con el prompt para cada una de las diferentes secciones del informe.
La Tabla \ref{tb4} despliega las versiones finales de los prompts, cada uno consiste en frases cortas, instrucciones claras y específicas para guiar las respuestas del modelo.


\begin{table}[!htb]
    \centering
    \caption{Prompts formulados para cada sección del informe geotécnico.}
    \label{tb4}
\scalebox{0.795}{    
    \begin{tabular}{p{0.5cm} p{12.5cm} p{5.5cm}} 
        \toprule
        \bfseries Sec. & \bfseries Prompt & \bfseries Resultados \\     
        \midrule
        \multirow{4}{*}{\rotatebox{90}{Objetivos}} & ``Eres un redactor técnico. Tu tarea es redactar los objetivos para un informe geológico basándote únicamente en el siguiente título del proyecto escribe sólo el objetivo general y dos objetivos específicos, en texto plano, sin encabezados, sin etiquetas, sin asteriscos ni formatos markdown. El texto debe ser claro, técnico y académico, y no debe repetir literalmente el título.'' & Se generan: 1 objetivo general y 2 objetivos específicos, a partir del título del proyecto. \\
        \midrule
        \multirow{10}{*}{\rotatebox{90}{Introducción}} & ``Eres un experto geólogo con 20 años de experiencia en caracterización de afloramientos rocosos. A partir de los datos técnicos de varios afloramientos, redacta una introducción técnica detallada que integre toda la información de manera coherente. El texto debe ser académico pero claro, con párrafos bien estructurados de aproximadamente 100 palabras cada uno. Evita listar datos crudos y en su lugar realiza síntesis interpretativa con textos. Usa conectores lógicos y mantén un flujo narrativo. No incluyas títulos ni encabezados. Datos de los afloramientos:'' & Una introducción preliminar en párrafos de aproximadamente 100 palabras cada uno, sin incorporar los datos ingresados por el usuario. \\
        & ``Con esta información, redacta una introducción técnica detallada que: Presente el contexto geológico general. Describa las características principales de los afloramientos. Analice los tipos de rocas encontradas y sus propiedades. Evalúe la calidad general de los macizos rocosos. Integre las observaciones de campo con interpretaciones geológicas. El texto debe fluir naturalmente, evitando listas o enumeraciones. Usa lenguaje técnico preciso pero accesible.'' & El LLM modifica la introducción preliminar integrando los datos ingresados por el usuario. \\
        \midrule
        \multirow{6}{*}{\rotatebox{90}{Afloramiento}} & ``Redacta un párrafo técnico y coherente (alrededor de 100 palabras) que describa el afloramiento geológico visible en la imagen, desde una perspectiva geotécnica. Analiza la estructura general del macizo rocoso, el tipo y densidad de fracturamiento, la orientación de las discontinuidades (si es inferible), y la estabilidad del talud, así como su calidad (buena, moderada, mala) y cualquier rasgo que implique riesgos geotécnicos como deslizamientos, caídas de bloques o colapsos. Incluye también observaciones sobre propiedades mecánicas visibles como compacidad, meteorización superficial o heterogeneidad del material. Evita listas; redacta todo como un párrafo fluido y técnico.'' & Detalle del afloramiento, incluye la estructura y calidad del macizo rocoso, el patrón de fracturamiento y discontinuidades, su estabilidad y posibles riesgos geotécnicos, junto con observaciones sobre su compacidad, meteorización y homogeneidad del material. \\
        \midrule
        \multirow{5}{*}{\rotatebox{90}{Muestra de mano}} & ``Redacta un párrafo técnico y detallado (alrededor de 100 palabras) que describa la muestra de roca visible en la imagen, desde un enfoque geotécnico. Describe el tipo de roca (si es identificable), textura, grado de compactación, fracturamiento o fisuras presentes, resistencia mecánica esperada, meteorización visible y cualquier debilidad estructural o heterogeneidad relevante. Enfócate en características que puedan afectar el comportamiento de la roca en campo, como la estabilidad o la deformabilidad. El texto debe ser fluido, sin enumeraciones, con lenguaje técnico claro y redactado como un párrafo continuo.'' & Descripción en un párrafo enfocada en las propiedades que pueden influir en su comportamiento en campo, especialmente en términos de estabilidad. \\
        \midrule
        \multirow{4}{*}{\rotatebox{90}{Esclerómetro}} & ``Eres un experto geotécnico. Con base en los siguientes resultados del esclerómetro Schmidt: Método, HR promedio 10 mayores, HR mediana 10 mayores, Peso específico, UCS media, UCS mediana, Módulo de Young (E). Redacta una interpretación clara y profesional que explique qué indican estos resultados sobre la calidad y resistencia de la roca, posibles aplicaciones y recomendaciones. El texto debe ser técnico, coherente y claro, sin repetir datos textualmente.'' & Descripción de la calidad de la roca, interpretando su resistencia, dureza y comportamiento mecánico, junto con recomendaciones de uso en obras civiles y ensayos complementarios. \\
        \midrule
        \multirow{11}{*}{\rotatebox{90}{Discusión}} & ``Como geólogo experto, analiza los siguientes datos de afloramientos rocosos y genera una discusión técnica detallada. El texto debe ser interpretativo, con análisis de patrones y relaciones geológicas, incluyendo evaluación de calidad del macizo rocoso mediante los índices RMR y SMR. Incluye inferencias basadas en la evidencia y plantea hipótesis cuando sea pertinente. Estructura el contenido en 3-4 párrafos coherentes, sin títulos ni enumeraciones. Datos para el análisis:'' & Interpretación de la calidad y comportamiento del macizo rocoso con datos de afloramientos, analizando patrones geológicos y estructurales, evaluando los índices RMR y SMR para determinar la estabilidad y condiciones mecánicas. \\
        & ``Desarrolla una discusión técnica que integre los siguientes aspectos: Relación entre las características petrográficas, tipo de roca y su origen geológico. Evaluación de la calidad del macizo rocoso basada en los índices RMR y su variabilidad. Análisis de estabilidad de taludes mediante el índice SMR, considerando los factores de ajuste. Interpretación de los patrones estructurales observados y su influencia en la calidad del macizo. Correlaciones entre los diferentes afloramientos y su significado geológico. Implicaciones para la ingeniería geológica y posibles riesgos identificados. El texto debe ser fluido, técnico, pero claro, con un enfoque interpretativo. Utiliza conectores adecuados para mantener la coherencia y evita listados numerados. Destaca las relaciones más relevantes entre los parámetros analizados.'' & Relaciona detalladamente las características petrográficas y el tipo de roca con su origen geológico, evaluando la calidad del macizo a través del índice RMR, patrones estructurales y su impacto en la calidad del macizo, estableciendo correlaciones entre diferentes afloramientos para comprender su contexto geológico. \\
        \midrule
        \multirow{4}{*}{\rotatebox{90}{Conclusiones}} & ``Como geólogo senior, sintetiza las principales conclusiones técnicas derivadas del estudio de los siguientes afloramientos rocosos. Presenta entre 4 y 6 conclusiones numeradas, cada una como un párrafo breve (2-3 oraciones). Las conclusiones deben ser específicas, basadas en evidencia y relevantes para la caracterización geotécnica.'' & Listado estructurado de 4 a 6 conclusiones cortas enfocadas en las propiedades mecánicas de las rocas y las implicaciones geotécnicas, fundamentadas en evidencia de los datos analizados. \\        
        \bottomrule
    \end{tabular}
    }
\end{table}

El diseño de cada prompt requiere un proceso de creación, refinación, adaptación y evaluación:

\begin{enumerate}
    \item \textit{Creación}: cada prompt empieza con una breve instrucción que indica de manera general el contenido requerido por cada sección. Se prueba con el MLLM y la respuesta generada es contrastada con la esperada para refinar el prompt.
    \item \textit{Refinación}: cuando la descripción generada es demasiado genérica, omite elementos de interés y el resultado carece de claridad, coherencia y precisión, la instrucción se divide en frases más cortas y concisas, cada una funcionando como una instrucción específica para el MLLM. Este paso implica mejorar la redacción, incorporar términos especializados y delimitar con mayor precisión los aspectos que debían ser abordados.    
    \item \textit{Adaptación}: las frases se detallan progresivamente con el objetivo de tener más insumos para la descripción. Se incorporan nuevas características y parámetros a partir de datos ingresados por el usuario, como características específicas de las rocas, la calidad de la roca en afloramientos y medidas estructurales, así como las mismas descripciones generadas por la IA para rocas y afloramientos que constituyen entradas de texto como parte del prompt para las secciones de introducción y discusión. 
    \item \textit{Evaluación}: Se valida el resultado comparándolo con la descripción del experto. Los pasos 2 al 4 se realizan de manera iterativa hasta obtener una respuesta adecuada.
\end{enumerate}

En nuestra evaluación, una respuesta adecuada es entendida como aquella que presenta claridad, coherencia y precisión técnica en la caracterización del afloramiento.
Esto incluye la correcta identificación de la litología, estructuras geológicas y condiciones geotécnicas del macizo rocoso.
Si la respuesta no cumplía estos criterios, la afinación del prompt estructurado ajusta el texto de entrada para orientar mejor al modelo.
Esto permitió obtener resultados más consistentes y técnicamente válidos, contribuyendo a la generación automatizada de informes geotécnicos con mayor nivel de detalle y fiabilidad.


\subsection{Evaluación}
\begin{figure}[!htb]
    \centering
    \includegraphics[width=0.9\linewidth]{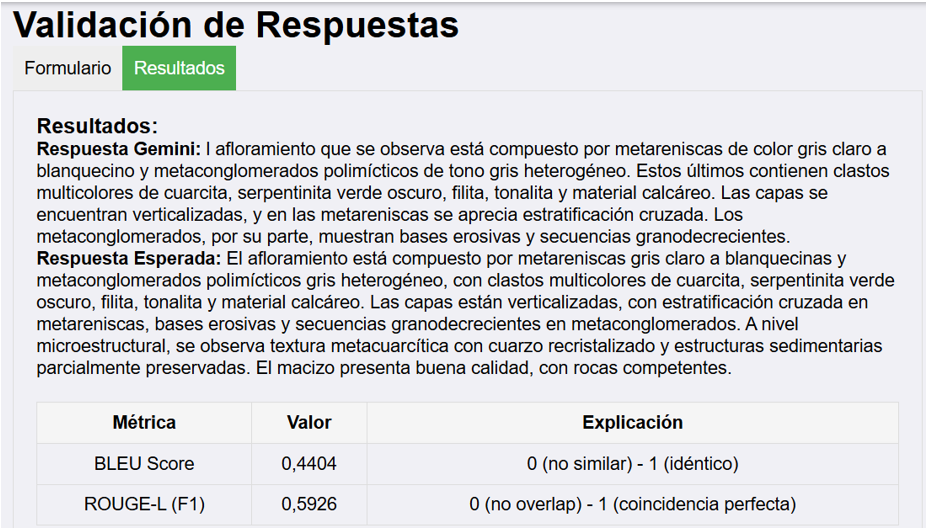}
    \caption{Comparación automática entre la descripción generada por Gemini 1.5 Flash y la descripción técnica, evaluadas con BLEU y ROUGE-L.}
    \label{fig:metricasbarplot1}
\end{figure}

La calidad de las descripciones geológicas generadas por Gemini 1.5 Flash fue evaluada automáticamente mediante la comparación con descripciones técnicas obtenidas de la plataforma V3Geo.
Se utilizaron imágenes de 30 afloramientos de rocas sedimentarias, metamórficas e ígneas (intrusivas y extrusivas).
Para cuantificar la similitud entre los textos generados por la IA y los originales, se emplearon las métricas BLEU y ROUGE-L (F1), ampliamente utilizadas en procesamiento de lenguaje natural para medir precisión y cobertura textual, respectivamente.
Un ejemplo de evaluación se muestra en la Figura \ref{fig:metricasbarplot1}.

\begin{figure}[!htb]
    \centering
    \includegraphics[width=0.85\linewidth]{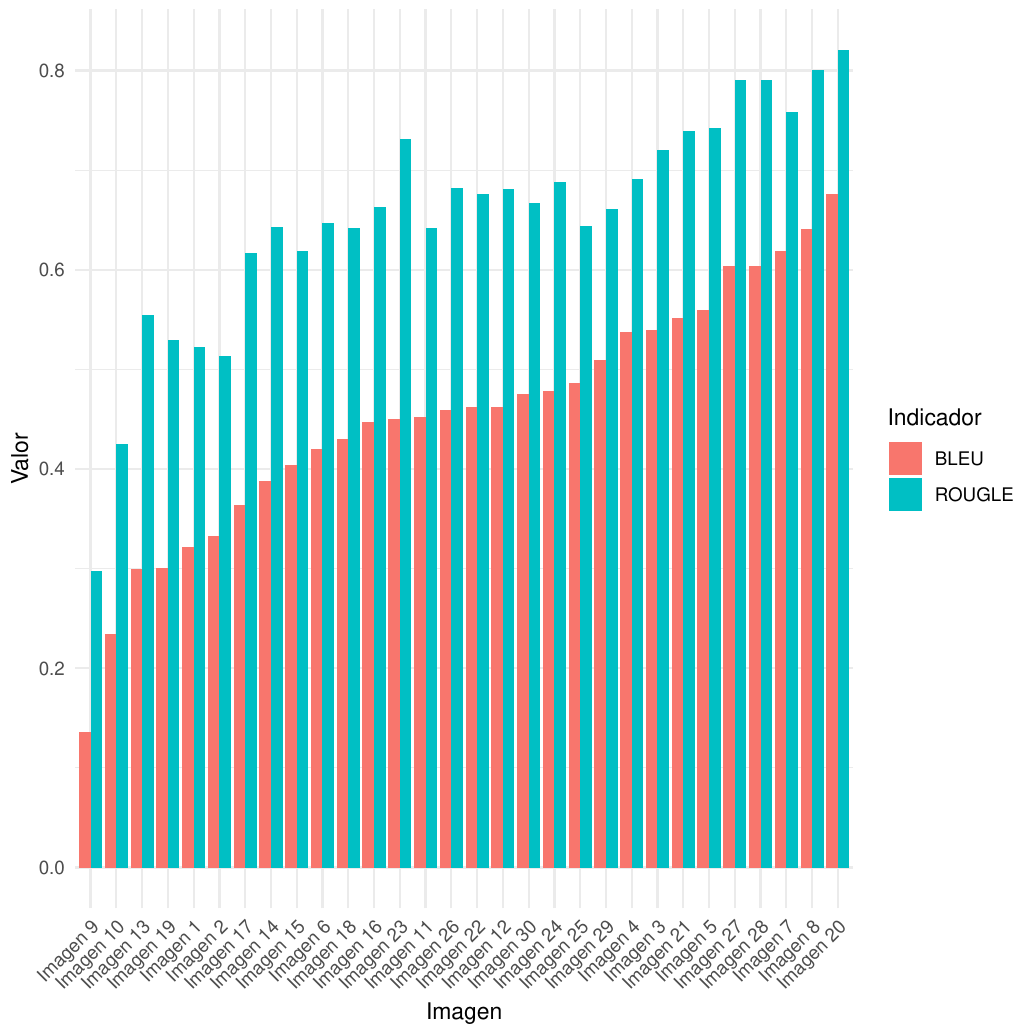}
    \caption{Evaluación automática de descripciones geológicas en 30 afloramientos de macizos rocosos.}
    \label{fig:metricasbarplot}
\end{figure}

Para facilitar el proceso de evaluación, se implementó una interfaz web desarrollada en \textit{Django}, donde se ingresaron ambas descripciones por afloramiento en un formulario que permite comparar directamente la respuesta del modelo con la respuesta esperada.
Al activar el botón de evaluar, el sistema ejecuta el análisis automático y calcula los puntajes de similitud, almacenándolos para su visualización en la pestaña de resultados. Estos se representan gráficamente en la Figura \ref{fig:metricasbarplot}, la cual permite observar el desempeño del modelo según el tipo de roca y validar la coherencia técnica de las descripciones generadas.

La métrica ROUGE-L muestra valores superiores a BLEU, lo cual refleja su mayor capacidad para capturar similitudes estructurales y secuencias largas entre los textos evaluados. Esta diferencia constante sugiere que ROUGE-L favorece evaluaciones más generosas sobre la calidad textual, mientras que BLEU proporciona un enfoque más estricto centrado en la precisión. El comportamiento ascendente en ambas métricas a lo largo de las imágenes indica una mejora progresiva en los textos analizados. 

\begin{figure}[!htb]
    \centering
    \subfloat[Correlación BLEU y ROUGE-L]{\includegraphics[width=0.33\textwidth]{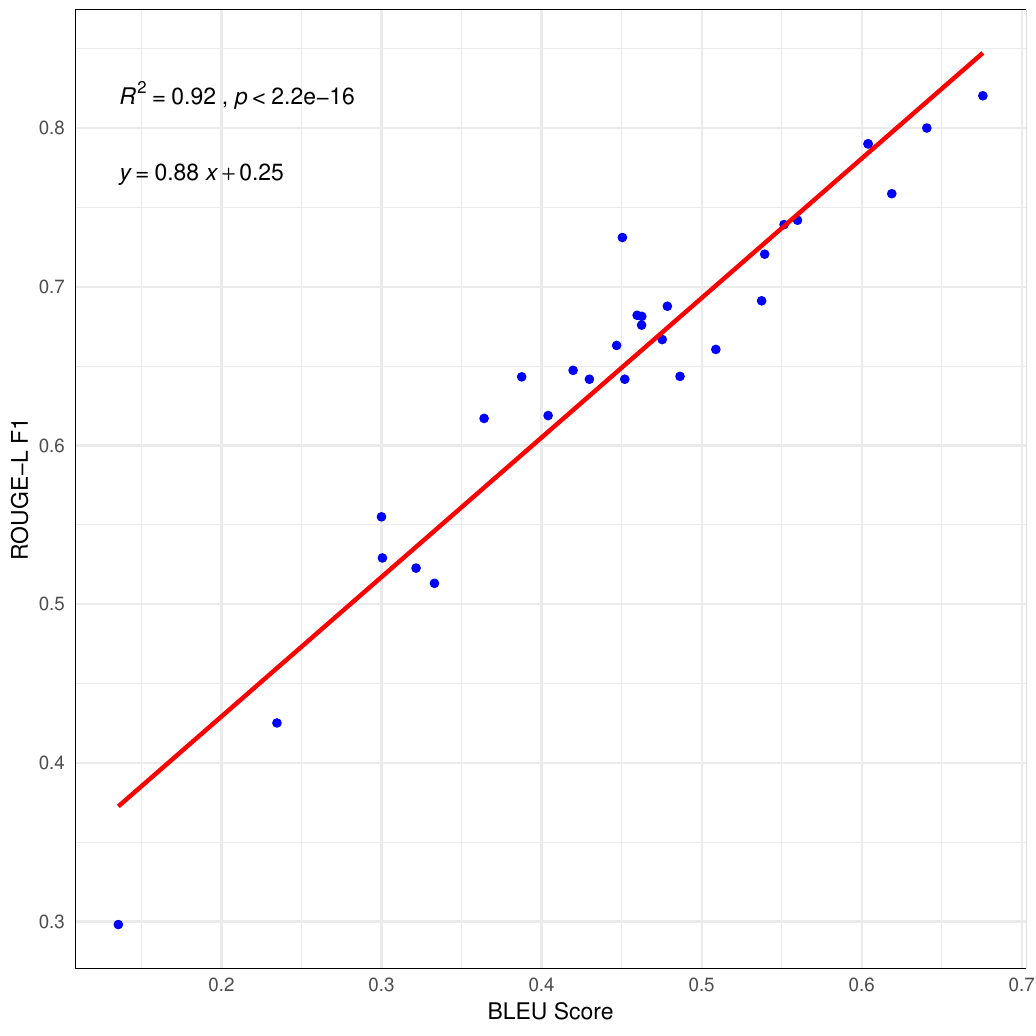}
        \label{fig:metricaA}
    }
    \subfloat[Histograma BLEU]{\includegraphics[width=0.33\textwidth]{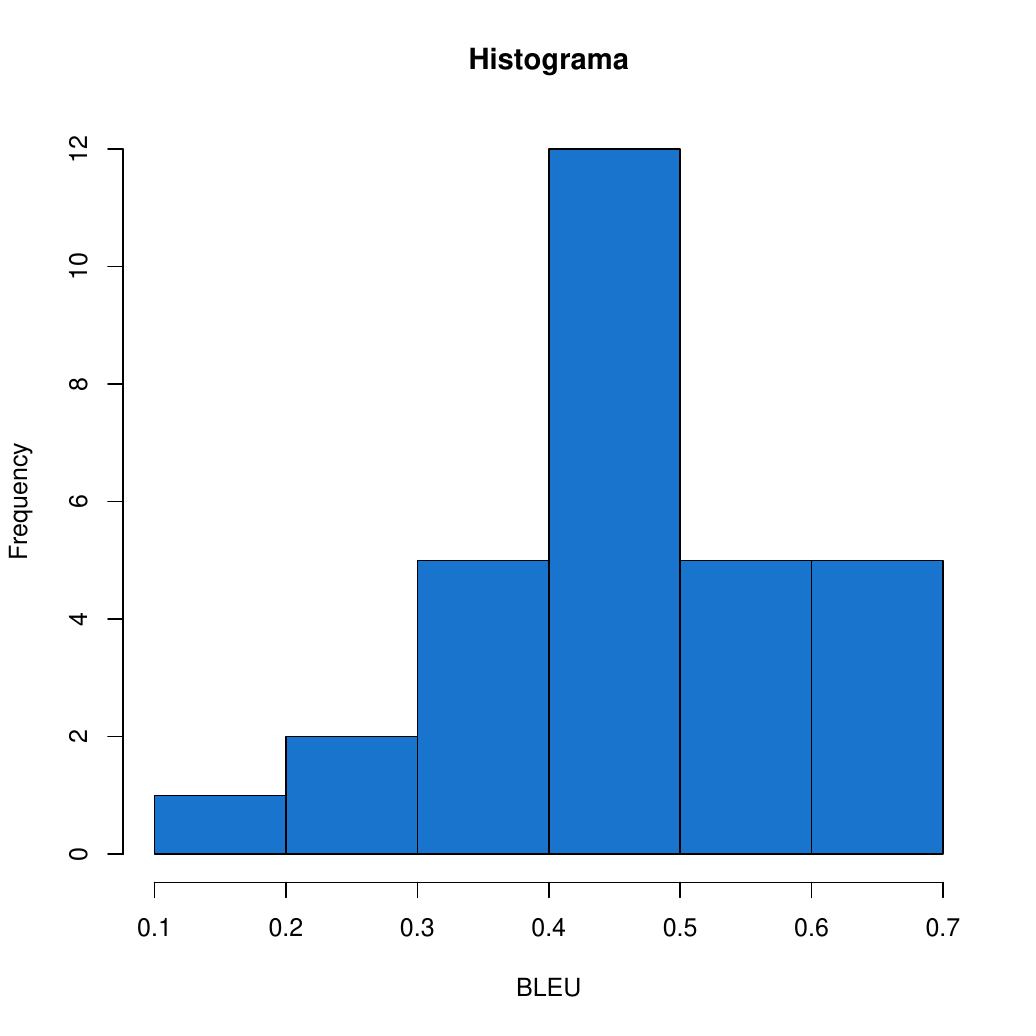}
        \label{fig:metricaB}
    }
    \subfloat[Histograma ROUGE-L]{\includegraphics[width=0.33\textwidth]{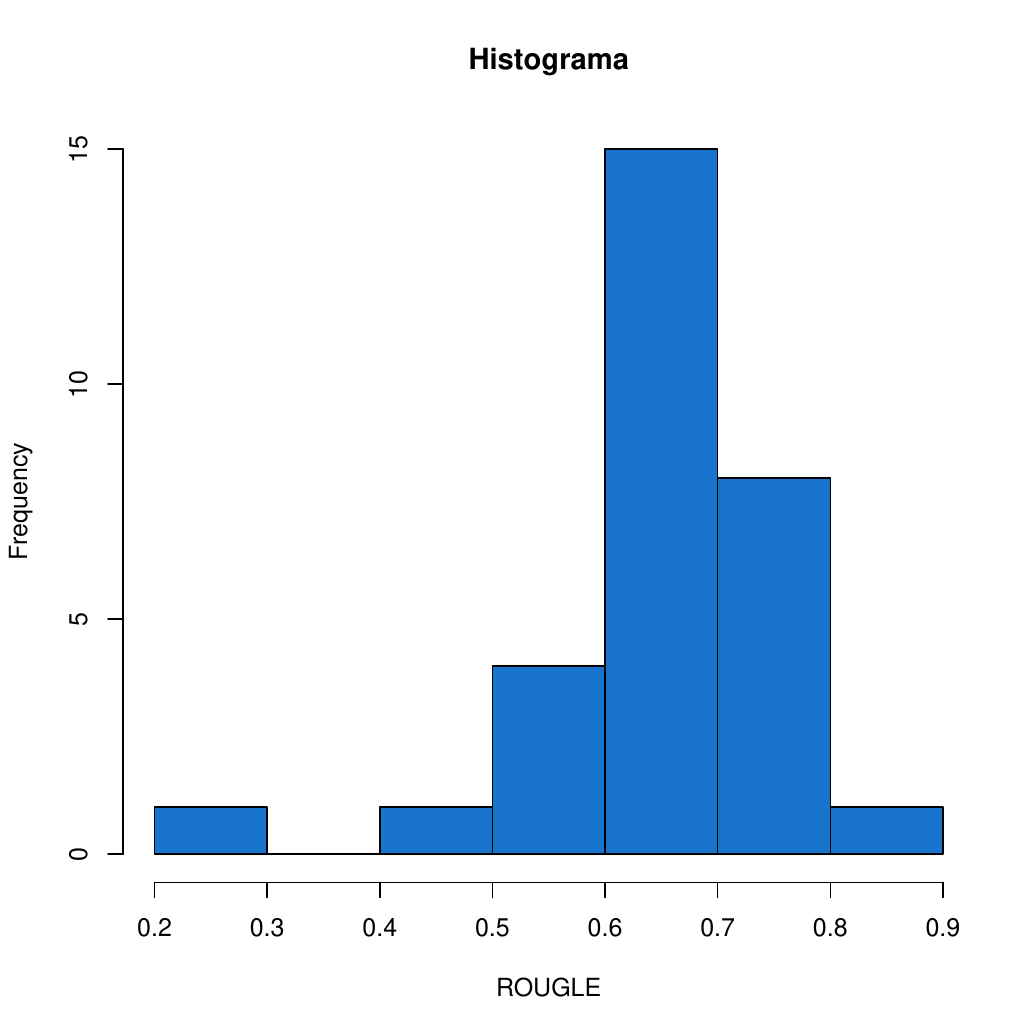}
        \label{fig:metricaC}
    }
    \caption{Valores de las métricas utilizadas en la evaluación.}
    \label{fig:metricas}
\end{figure}

BLEU mide la precisión y ROUGE-L evalúa la coincidencia secuencial. Las gráficas estadísticas más representativas de estas métricas se presentan en la Figura \ref{fig:metricas}.
La Figura \ref{fig:metricaA} indica una alta correlación lineal ($R^2= $ 0.92) entre BLEU y ROUGE-L, lo que significa que una métrica puede usarse como un indicador de la otra, ya que sus puntuaciones tienden a moverse en la misma dirección y de manera proporcional.
Esto sugiere que si el modelo es preciso, también será capaz de generar los aspectos clave del texto de referencia.
Las Figuras \ref{fig:metricaB} y \ref{fig:metricaC} muestran la distribución de valores de las métricas BLEU y ROUGE-L, respectivamente.
BLEU presenta una concentración de frecuencias entre 0.4 y 0.5, con una media de 0.455 y una mediana de 0.461, valores muy cercanos que sugieren una distribución casi simétrica, posiblemente normal.
Por otro lado, ROUGE-L tiene su clase modal entre 0.6 y 0.7, con una media de 0.653 y una mediana de 0.665, lo que indica una ligera asimetría positiva asociada a valores altos atípicos.
Esta comparación destaca cómo BLEU se comporta con mayor estabilidad estadística, mientras que ROUGE-L muestra mayor dispersión. 

Estos resultados, a partir del uso de métricas objetivas, sugieren que Gemini 1.5 Flash ofrece descripciones coherentes y técnicamente alineadas con las originales, siendo especialmente importantes para nuestra aplicación donde la integridad descriptiva es crucial.
La experimentación realizada ha demostrado que el proceso de ingeniería de prompt, basado en instrucciones estructuradas y específicas, es eficaz para alcanzar las respuestas deseadas, convirtiéndolo en una base sólida para futuras integraciones de IA en procesos de reporte geotécnico y estructural.

\section{GeoReportIA: Generador de informes geotécnicos}
\label{sec:app}
Es común que los proyectos de IA se limiten a un entorno local o de manera restringida en la nube.
Integrar un modelo en una aplicación y pasar a la fase de producción implica un proceso complejo y demanda una inversión significativa de recursos y tiempo \cite{Paleyes2021Challenges}.
Sin embargo, el objetivo principal de este trabajo es la creación de una herramienta funcional tanto para el ámbito académico como profesional.

La tecnología web es la plataforma más adecuada para crear aplicaciones gratuitas, de acceso libre y que sean fáciles de usar e intuitivas.
Un estudio reciente de Vidal-Silva et al. (2021) destaca la eficacia de Django, un marco de trabajo (framework) web basado en Python, para el rápido desarrollo de sistemas de información web en entornos académicos.
Esta es la herramienta empleada para la creación de la aplicación web de generación automática de informes geotécnicos, ya que facilita el desarrollo ágil de aplicaciones mediante una arquitectura clara y eficiente \cite{djangoproject2025}.


\subsection{Plataforma computacional} 

El proyecto se llevó a cabo en un equipo con un procesador AMD Ryzen 3 3300 U de 2.10 GHz, 8 GB de RAM, tarjeta gráfica AMD Radeon Vega 6 de 2 GB y 1.38 TB de almacenamiento (SSD + HDD).
En cuanto al software, Windows 11 como sistema operativo, el lenguaje de programación Python 3.13.0 se utilizó para la configuración de Django, las funciones de cálculo y el enlace con el modelo Gemini 1.5 Flash mediante la API Key; mientras que
JavaScript 1.8.0 para la interactividad, funcionalidad y generación de documentos PDF, junto con HTML y CSS para la estructura, el estilo y la visualización en el entorno web.

Para el desarrollo del código se utilizó el editor \textit{Visual Studio Code}, empleando un entorno virtual que asegura la portabilidad del proyecto y evita conflictos entre versiones de librerías.
Las librerías usadas son múltiples, un listado completo se encuentra en el archivo 
\href{https://github.com/Chris-2801/Informes_Geotecnicos_IA/blob/main/requirements.txt}{\textit{requirements.txt}}. Entre ellas, destacan Django para el desarrollo web; \textit{google-ai-generativelanguage} y \textit{google-api-core} para enlazar con Gemini 1.5 Flash; \textit{gunicorn} para la etapa de producción con el servidor web; y \textit{scikit-learn}, \textit{safetensors} y \textit{torch} para el correcto funcionamiento de nuestro modelo.


\subsection{Arquitectura del sistema}
El diseño y la implementación de una aplicación web es un proyecto no trivial.
En lugar de comenzar desde cero, apoyarse en un \textit{framework} facilita y acelera drásticamente el proceso al seguir un patrón de diseño y las mejores prácticas.
Django es un conocido framework de Python que proporciona una estructura y funcionalidades básicas como la gestión de URLs, la interacción con bases de datos y la creación de plantillas.
Un proyecto en Django se organiza bajo un esquema MVT (Modelos para estructura de datos, Vistas para la lógica y Templates para la presentación), separando la funcionalidad (lógica) de la presentación para crear un flujo claro y organizado.

En esencia, la organización de nuestra aplicación web sigue la estructura clásica de un proyecto en Django. La distribución de los directorios, subdirectorios y archivos es la siguiente:

\begin{verbatim}
GeoReportIA/
|-- ProyectowebApp/
|   |-- models.py           # Define la estructura de la base de datos  
|   |-- urls.py             # Define las rutas (URLs) y las asocia a una vista específica             
|   |-- utils.py            # Funciones auxiliares utilizadas por las vistas 
|   |-- views.py            # Cerebro de la aplicación             
|   |-- templates/          # Aquí se guardan los archivos HTML  
|   |   |-- ProyectowebApp/
|   |       |-- index.html
|   |       |-- page1.html
|   |       |-- page2.html
|   |       |-- ...  
\end{verbatim}

\begin{figure}[!htb]
    \centering
    \includegraphics[width=0.9\linewidth]{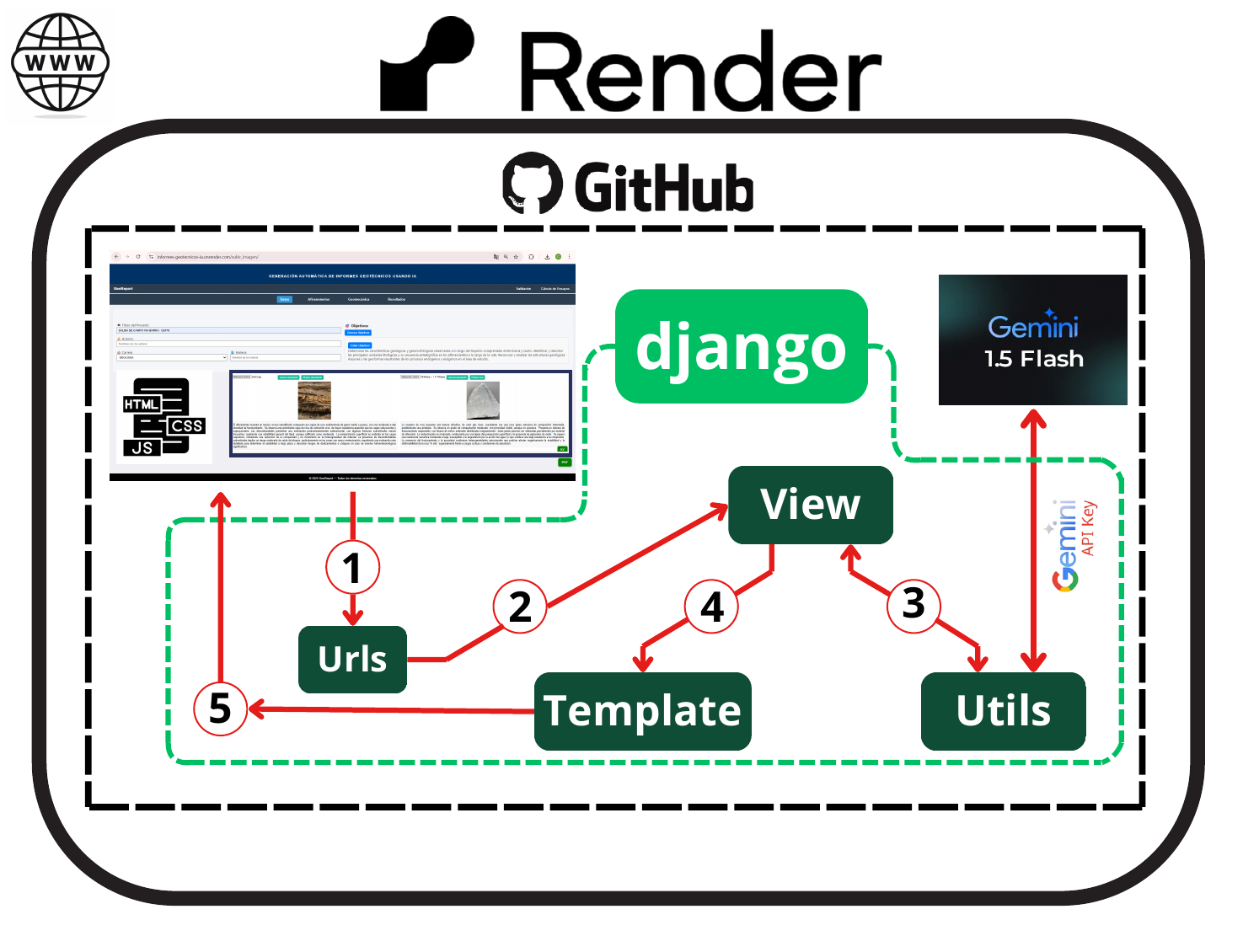}
    \caption{Esquema de la arquitectura web de la aplicación: componentes e interacción.}
    \label{Arquitectura}
\end{figure}

Dentro de cada aplicación, se encuentran los componentes clave como View, Template, Utils y Urls.
El diagrama de la Figura \ref{Arquitectura} esquematiza una aplicación web basada en Django, ilustrando la interacción entre sus elementos y con el resto del sistema. 


El funcionamiento se interpreta como un flujo de solicitud-respuesta a través de los siguientes pasos:

\begin{enumerate}
    \item El flujo comienza cuando un usuario realiza una solicitud a la aplicación a través del navegador.
    \item Django recibe esta solicitud y la dirige al enrutador de URLs, que la compara con el archivo de configuración de URLs de la aplicación para encontrar la vista asociada. Esta vista, a su vez, contiene el código que debe ejecutarse para procesar la solicitud.
    
    \item La vista contiene la lógica central y las funciones de la aplicación. Se encarga de procesar las solicitudes de los usuarios y decidir qué respuesta generar. Aquí puede invocar funciones adicionales (utils) o conectarse con la base de datos (aunque en el diagrama no aparece explícitamente el Model).
    \item Una vez que la vista ha procesado la lógica, utiliza un template y le ``inyecta'' la información que procesó, generando la respuesta final. Esta plantilla define la estructura y el diseño de la página web utilizando HTML, CSS y JavaScript para su presentación.
    \item Por último, la vista envía esta plantilla en forma de página web al navegador del usuario como respuesta final.
    

\end{enumerate}

Por ejemplo, el usuario introduce la URL de inicio de la aplicación o hace clic en el enlace para el ingreso de datos, como cargar una imagen de afloramiento rocoso.
Django identifica la URL asociada en el archivo urls.py y redirige la petición hacia la vista correspondiente.
La vista recibe el archivo subido por el usuario y llama a una función de utils.py que se encargue de invocar a Gemini 1.5 Flash a través de la clave API.
Luego, la imagen es pasada al modelo junto con el prompt de la sección respectiva del informe geotécnico.    
El modelo genera una descripción textual de la imagen y se devuelve a la vista.
La vista inserta el resultado en un template HTML, junto con la imagen cargada. El usuario ve en el navegador la imagen original junto con la descripción generada por el modelo.

\subsection{Despliegue de la aplicación}
\label{subsec:despliegue}

La creación de la aplicación web sigue una estrategia de desarrollo en un entorno local y posterior despliegue en una plataforma web en la nube.
Por ende, la arquitectura completa del sistema involucra otros componentes como el servidor web, la base de datos y, en este caso, la integración con el MLLM y otros servicios.

En la computadora local, la aplicación aprovecha la dirección IP 127.0.0.1, también conocida como \textit{localhost}, y se aloja en el puerto 8000, accesible con la URL: \texttt{http://127.0.0.1:8000/}.
El modelo de Gemini se ejecutó directamente en el localhost en un template de prueba diseñado específicamente para mostrar la respuesta del modelo utilizando los prompts.

El diseño del formato del informe fue realizado en conjunto con la prueba del MLLM para garantizar que las respuestas generadas y mostradas en el template se impriman correctamente en formato PDF.
Una vez terminada la ingeniería del prompt para todo el informe, se procedió a diseñar los templates definitivos en donde se mostraría la respuesta de la IA.
Para esto se diseñó un template configurado para cuatro secciones: 1) descripción de rocas y afloramientos; 2) generación de las secciones de introducción, discusión y conclusiones; 3) los objetivos y los datos de la portada; y 4) clasificaciones geomecánicas.

Una vez probado y validado de manera local, el proyecto se sube al repositorio de \textit{GitHub}, el cual soporta control de versiones y permite publicar el código fuente en el siguiente enlace: \href{https://github.com/Chris-2801/Informes_Geotecnicos_IA.git}{\textcolor{blue}{GitHub}}.
La disponibilidad de la aplicación en línea requiere un servidor de alojamiento y ejecución.
\textit{Render}\footnote{\url{https://render.com/}} es una excelente opción para pasar una aplicación Django de un entorno de desarrollo local a un entorno en la nube, ya que simplifica significativamente el proceso de despliegue.
Esta plataforma en la nube automatiza muchas de las tareas complejas del despliegue, como la configuración del servidor y la conexión con el repositorio de código.

Antes de subirlo a la nube, se debe preparar el proyecto para producción.
Esto incluye ajustes en el archivo \textit{settings.py} de Django y la configuración de una base de datos de producción.
Una vez desplegada, la aplicación web está accesible a través de una URL pública: \href{https://informes-geotecnicos-ia.onrender.com/subir_imagen/}{\textcolor{blue}{GeoReportIA}}.
Esta herramienta puede ser aprovechada desde una computadora de escritorio, así como dispositivos móviles\footnote{La aplicación puede demorar pocos minutos en cargar por primera vez, dadas las limitaciones de la versión gratuita de Render}.

\subsection{Uso del sistema}
\label{subsec:uso}

\begin{figure}[!htb]
    \centering
    \includegraphics[width=1\linewidth]{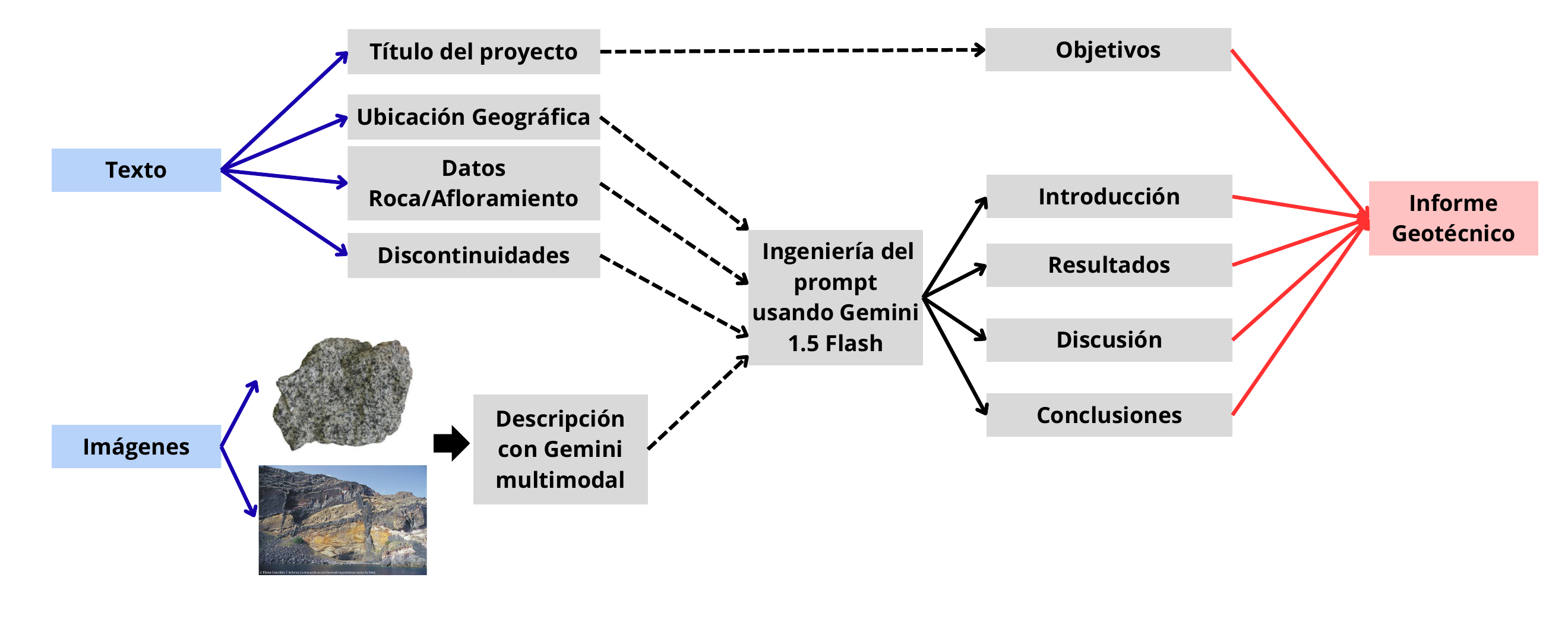}
    \caption{Esquema del proceso de generación del informe geotécnico.}
    \label{arquitectura_solucion}
\end{figure}

Cuando un usuario interactúa con la interfaz de la aplicación web, esta captura los datos ingresados. Luego, la aplicación combina esta información con los prompts definidos para que el MLLM los procese. Como resultado, el MLLM genera y devuelve un informe geotécnico con contenido automatizado en cada una de sus secciones.
Este flujo de entrada, proceso y salida se esquematiza en la Figura \ref{arquitectura_solucion}.

Existen dos grandes categorías de entrada de datos para la aplicación: texto e imágenes.
La entrada de texto consiste en varias secciones editables manualmente por el usuario.
Incluye información fundamental como el título del proyecto, la ubicación geográfica, los datos acerca de la roca o afloramiento y las discontinuidades encontradas en campo.

En el caso de afloramientos, el usuario ingresa información relevante recopilada en campo, como coordenadas, tipo de roca, matriz, textura, entre otros datos.
Asimismo, se pueden registrar los datos de las familias estructurales identificadas en el afloramiento, los cuales se utilizan posteriormente para generar la representación gráfica en un estereograma.

La sección de rocas y afloramientos está organizada en bloques; cada uno de ellos representa un punto de muestreo, y cada punto permite generar descripciones de imágenes del afloramiento y muestra de mano. Además, dispone de un apartado estructural y otro para descripciones que para la IA se le dificultan visualmente, como el tipo y nombre de roca, la calidad, la ubicación geográfica y características más específicas como la mineralogía y la presencia de matriz. 

La entrada de imágenes permite al usuario cargar fotografías tanto de afloramientos como de muestras de mano, que son interpretadas visualmente por el modelo extrayendo características relevantes, a partir de las cuales se genera automáticamente la descripción detallada correspondiente.


Tanto los datos ingresados por el usuario como las descripciones generadas a partir de las imágenes fluyen hacia el centro del sistema, donde se encuentra el modelo Gemini 1.5 Flash, potenciado mediante ingeniería de prompts.
Este modelo procesa toda la información y la organiza en las secciones que conforman el informe: introducción, resultados, discusión y conclusiones. Es importante destacar que los objetivos del informe de campo se generan de forma directa a partir del título del proyecto. 

Las secciones de introducción, discusión y conclusiones están compuestas por párrafos de no más de 100 palabras; estos párrafos pueden ser editados por el usuario.
Una vez revisado y completado el contenido, el informe geotécnico final puede ser descargado en formato PDF.
Este producto es completamente automatizado, abordando la problemática de la redacción manual, la subjetividad y la complejidad de consolidación de la información. Un ejemplo del informe generado se muestra en el siguiente enlace: \href{https://drive.google.com/file/d/1OjNvchAZEpf8avx38mvAPqTrdXotMELu/view?usp=drive_link}{\textcolor{blue}{Ejemplo de informe}}.

\section{Conclusión}
\label{sec:conclusion}

Este trabajo presentó el desarrollo de GeoReportIA, un sistema web de generación automática de informes geotécnicos para macizos rocosos, el cual procesa datos de entrada del usuario (imágenes, ubicación y datos técnicos) y genera de manera autónoma un informe completo y coherente.

Aprovechando el poder de un modelo de lenguaje multimodal como Gemini 1.5 Flash y una cuidadosa ingeniería de prompts, el método propuesto evita la necesidad de un reentrenamiento costoso del modelo de IA, resultando en una optimización significativa de recursos y tiempo.

La recopilación de imágenes de afloramientos rocosos con sus respectivas descripciones en texto y el análisis de informes geotécnicos de campo
fueron las actividades fundamentales para la definición de la estructura del informe, la creación de prompts estructurados y específicos para cada sección del informe y la validación de las respuestas del MLLM.

La integración de tecnologías como Django, un repositorio de código como GitHub y un servidor de despliegue como Render, permitió el desarrollo de una aplicación gratuita y disponible en línea, proporcionando una herramienta valiosa para profesionales y estudiantes.
Este enfoque puede ser útil para el desarrollo de aplicaciones similares para la automatización de informes estándares y precisos en otras disciplinas de las Ciencias de la Tierra.

Si bien la arquitectura propuesta, impulsada y optimizada por la formulación de instrucciones para el MLLM, ofrece una solución eficaz, robusta y escalable, el trabajo futuro podría centrarse en la incorporación de modelos predictivos adicionales y la mejora del rendimiento para manejar un mayor volumen de datos de campo.

\bibliographystyle{unsrt}
\bibliography{references}

\clearpage
\setlength{\columnsep}{1.5cm}
\twocolumn 
\section*{Autores}

{\flushleft\textbf{Christofer Oswaldo Valencia Carranco}}

\begin{wrapfigure}{l}{15mm}
    \includegraphics[width=1in,height=2in,clip,keepaspectratio]{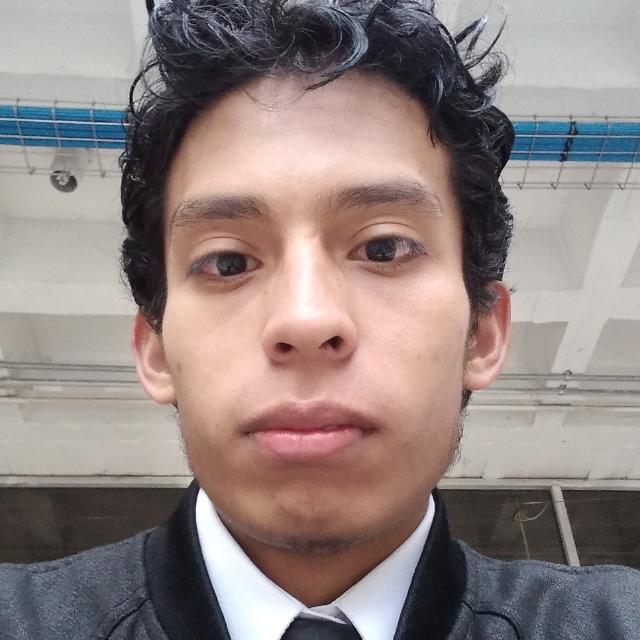}
\end{wrapfigure}\par

Nació en Ibarra, el 28 de enero de 2001. Cursa el 10mo semestre de la carrera de Geología en la Universidad Central del Ecuador. Realizó pasantías en: Instituto de Investigación Geológico y Energético (IIGE), en el área de Cartografía Regional; SIS ECU 911 en el área de gestión de datos; Ayudante de Cátedra de las materias de matemática y Dibujo técnico - CAD; Petroecuador EP en la Gerencia de Exploración y Producción del Activo Sacha. Tiene conocimientos en Mudlogging, SIG, Deep Learning, Python y desarrollo web. Actual tesista de Petroecuador EP en el activo Coca - Payamino Yuralpa.\\ 
\vspace{0.5cm}


{\flushleft\textbf{Alexis David Llumigusín Caiza}}

\begin{wrapfigure}{l}{15mm}
    \includegraphics[width=1in,height=2in,clip,keepaspectratio]{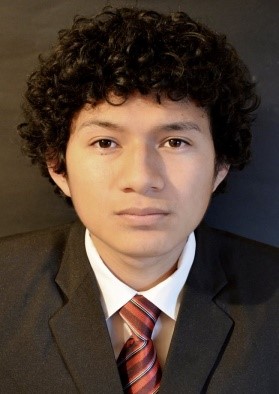}
\end{wrapfigure}\par

Nació en Machachi, el 7 de febrero de 1997. Cursa el 10mo semestre de la carrera de Geología en la Universidad Central del Ecuador. Realizó pasantías en el Instituto de Investigación Geológico y Energético (IIGE), en el área de Ocurrencias Minerales Metálicas en el Proyecto de Investigación Geológica y Disponibilidad de Ocurrencias Minerales en el Territorio Ecuatoriano. Posee experiencia en mapeo geológico, Sistemas de Información Geográfica (SIG) y Deep Learning.\\ 
\vspace{0.5cm}


{\flushleft\textbf{Abrahan Jorge Arias Larco}}

\begin{wrapfigure}{l}{15mm}
    \includegraphics[width=1in,height=2in,clip,keepaspectratio]{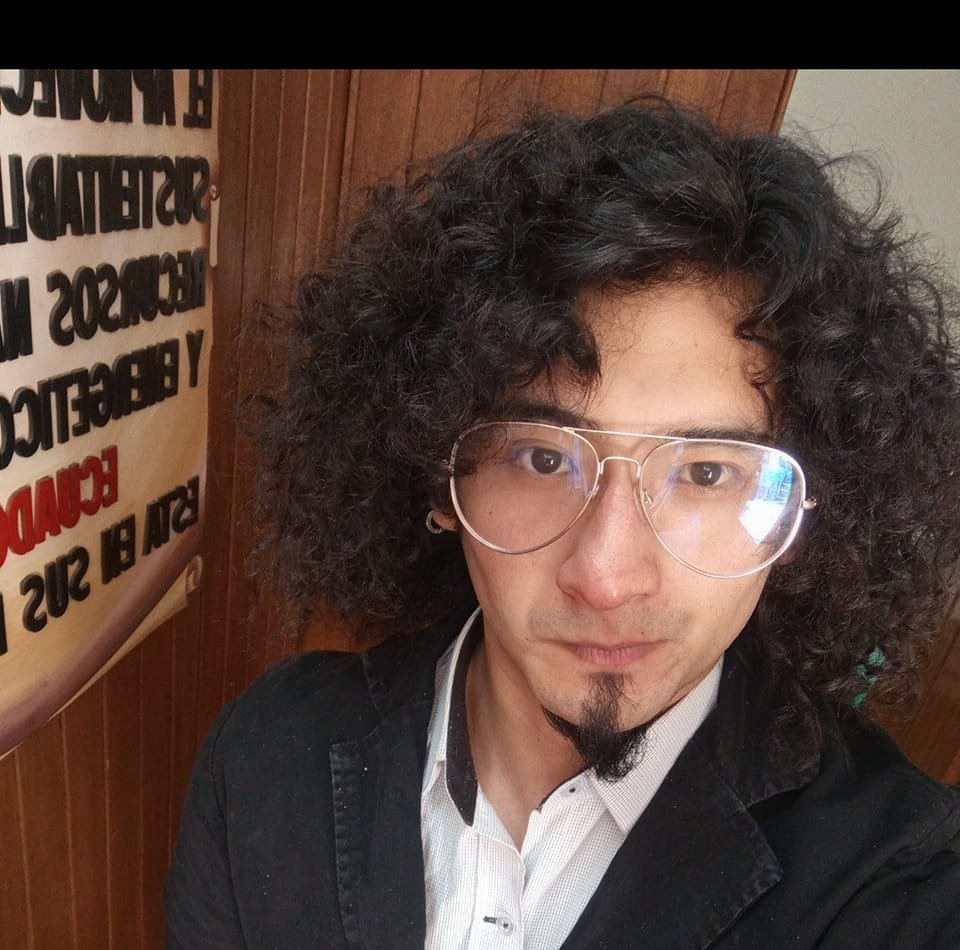}
\end{wrapfigure}\par

Nació en Quito, el 20 de julio de 1998. Cursa el 10mo semestre de la carrera de Geología en la Universidad Central del Ecuador. Realizó pasantías en el Instituto de Investigación Geológico y Energético (IIGE) y en el Ministerio del Medio Ambiente, Agua y Transición Ecológica en el área de laboratorio de metales pesados. Posee experiencia en geomática, muestreo de aguas y Deep Learning. \\

\vfill
\break


{\flushleft\textbf{Silvia Yessenia Alvarez Merino}}

\begin{wrapfigure}{l}{15mm}
    \includegraphics[width=1in,height=2in,clip,keepaspectratio]{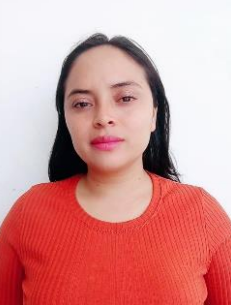}
\end{wrapfigure}\par

Nació en Loja, el 30 de octubre de 1995. Cursa el 10mo semestre de la carrera de Geología en la Universidad Central del Ecuador. Realizó pasantías en el Instituto de Investigación Geológico y Energético (IIGE), en el área de Cartografía Regional. Experiencia en cartografía geológica, Sistemas de Información Geográfica (SIG) y Deep Learning.\\ 
\vspace{0.5cm}

{\flushleft\textbf{Christian Iván Mejía Escobar}}

\begin{wrapfigure}{l}{15mm}
    \includegraphics[width=1in,height=1.25in,clip,keepaspectratio]{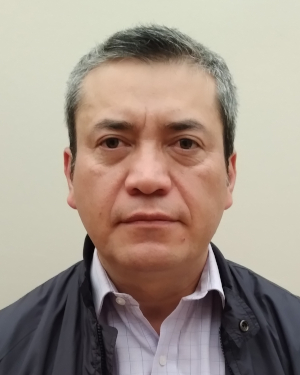}
\end{wrapfigure}\par

Profesor e investigador sobre Inteligencia Artificial en la Universidad Central del Ecuador. PhD en Informática de la Universidad de Alicante, España, 2023. Experiencia en proyectos de Machine Learning y Deep Learning, especialmente en aplicaciones de visión por computadora y procesamiento del lenguaje natural. \\

\end{document}